\documentclass[twocolumn,aps]{revtex4}
\usepackage{dcolumn}
\usepackage{bm}
\usepackage{amssymb}
\usepackage{amsmath}
\usepackage{epsf,amsmath,graphicx,verbatim}  

\begin{document}

\title{
Density functional theory and Molecular Dynamics Studies on Energetics and
Kinetics for Electro-Active Polymers: PVDF and P(VDF-TrFE)
}

\author{Haibin Su$^1$, Alejandro Strachan$^2$, and William A. Goddard, III$^1$}
\affiliation{
$^1$Materials and Process Simulation Center, Beckman Institute (139-74)\\
California Institute of Technology, Pasadena, California 91125\\
$^2$Theoretical Division, Los Alamos National Laboratory\\
Los Alamos, New Mexico 87545}
\date{\today}

\begin{abstract}
\noindent

We use first principles methods to study static and dynamical mechanical
properties of the ferroelectric polymer Poly(vinylidene fluoride) (PVDF) and 
its copolymer with trifluoro ethylene (TrFE). We use density functional theory 
[within the generalized gradient approximation (DFT-GGA)] to calculate structures 
and energetics for various crystalline phases for PVDF and P(VDF-TrFE). We find 
that the lowest energy phase for PVDF is a non-polar crystal with a combination 
of trans (T) and gauche (G) bonds; in the case of the copolymer the role of the 
extra (bulkier) F atoms is to stabilize T bonds. This leads to the higher 
crystallinity and piezoelectricity observed experimentally. Using the MSXX first 
principles-based force field (FF) with molecular dynamics (MD), we find that 
the energy barrier necessary to nucleate a kink (gauche pairs separated by trans bonds) in an all-T 
crystal is much lower (14.9 kcal/mol) in P(VDF-TrFE) copolymer than in PVDF (24.8 kcal/mol). This correlates 
with the observation that the polar phase of the copolymer exhibits a solid-solid 
a transition to a non-polar phase under heating while PVDF directly melts. We also 
studied the mobility of an interface between a polar and non-polar phases under
uniaxial stress; we find a lower threshold stress and a higher mobility in the 
copolymer as compared with PVDF. Finally, considering plastic deformation under applied shear, we find that the
chains for P(VDF-TrFE) have a very low resistance to sliding, particularly along the
chain direction. The atomistic characterization of these ``unit mechanisms" provides
essential input to mesoscopic or macroscopic models of electro-active polymers.

\end{abstract}

\maketitle

\section{Introduction}

Poly(vinylidene fluoride) (PVDF) and its copolymers with trifluoro ethylene (TrFE) 
exhibit excellent electro-mechanical properties such as ferroelectricity,
piezoelectricity, pyroelectricity, and nonlinear optical properties \cite{nalwa95}.
Ferroelectricity in PVDF and P(VDF-TrFE) random copolymers was demonstrated with
various experimental techniques [including X-Rays, IR, polarization-field hysteresis
loops, and the existence of a Curie point (only in copolymers)] in the late 70's
and early 80's \cite{furukawa89}. Very recently, Zhang and collaborators \cite{zhang98}
showed that it is possible to make use of electric field-induced phase transformations
between polar and non-polar phases in nano-structured (via electron irradiation)
P(VDF-TrFE) to obtain large electrostrictive strains ($\sim$ 5 $\%$) at high
frequencies (1000 KHz) and with good energy densities (comparable to the best
piezoceramics). Such materials are very attractive for many applications requiring
soft transducers, due to good acoustic impedance match with biological tissue 
and water.

Molecular dynamics (MD) has previously been used to study electro-mechanical properties 
of PVDF \cite{wag92, wag95} and P(VDF-TrFE) \cite{gjk94,gjk96,gjk98,abe01-3,abe01-4,
abe01-5} including the prediction of elastic, dielectric, and piezoelectric constants, 
and polar to non-polar phase transition. However, despite experimental and theoretical 
efforts and the technological importance of this class of materials, the fundamental 
molecular processes responsible for their macroscopic electromechanical properties are 
largely unknown. A quantitative characterization of such processes is necessary to build
predictive, first principles-based materials models and should be helpful for the design 
of materials with improved properties (such as actuators with higher strain, precision, 
frequency, and energy density).

In this paper we use atomistic modeling [DFT and first principles-based force fields
(FFs) with molecular dynamics (MD)] (described in section II) to study: i) structures
and energies of various crystal structures (section III); ii) the nucleation energy
associated with the polar to non-polar phase transformation and phase boundary
propagation as a function of applied external stress (section IV), and iii) chain
sliding under shear deformation (section V). Finally, conclusions are drawn in section 
VI. The comparison between the behavior of PVDF and P(VDF-TrFE) random copolymers allow 
us to quantify the effect of the trifluoro ethylene (TrFE) segments on various materials
properties.

\subsection{Crystal structures of PVDF}

\begin{figure*}
\includegraphics [scale=0.75]  {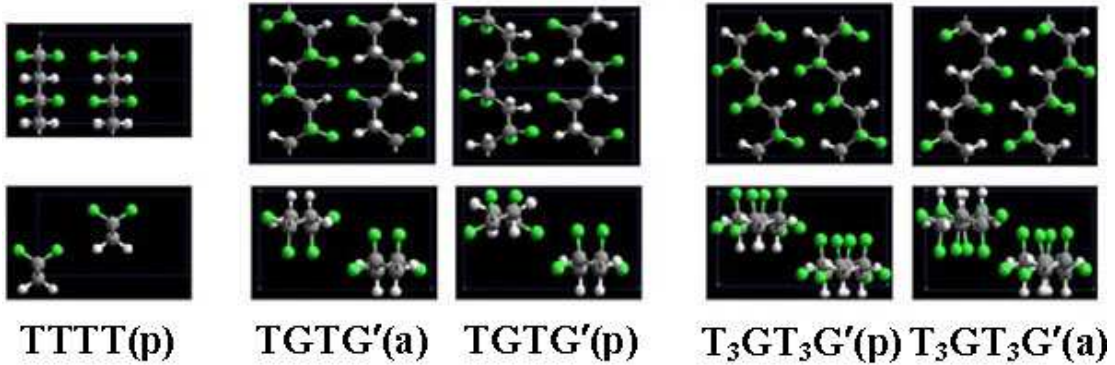}
\caption{\label{fig:1-stru} Crystal structures of various phases for PVDF. The polar
phases with parallel dipole moments are denoted by (p), while non-polar phases with
anti-parallel dipole moments are denoted by (a). The existence of the non-polar 
T$_3$G$_{ad}$ phase was conjectured by Lovinger \cite{lovinger81} and proved by Karasawa
and Goddard \cite{wag92}.}
\end{figure*}

Four crystalline polymorphs of PVDF are well characterized experimentally; they are 
generally referred to as I, II, III, and IV. Following \cite{wag92} we refer to these 
as follows: I = T$_{\rm p}$, II= TG$_{\rm ad}$, III = T$_3$G$_{\rm pu}$, and IV = 
TG$_{\rm pd}$ [see Fig.(\ref{fig:1-stru})]. In this notation we indicate the chain 
conformation with capital letters (T means all-T, TG means TGTG$'$ and T$_3$G indicates 
TTTGTTTG$'$) the subscripts p or a indicate polar phases with parallel dipoles and 
non-polar phases with anti-parallel dipole moments respectively; finally the subscripts 
u and d indicate up-up or up-down relative directions of adjacent chains \cite{wag92}. 
When PVDF is cooled from the melt, the TG$_{\rm ad}$ phase is obtained. The polymer 
chains contain T and G bonds in a TGT${\rm G}'$ sequence. This phase has a polar 
counterpart: the TG$_{\rm pd}$ phase. The main difference between them is the orientation 
of the dipole moments. In TG$_{\rm pd}$ they are parallel while in TG$_{\rm ad}$ they 
are anti-parallel. The TG$_{\rm ad}$ phase can be converted into TG$_{\rm pd}$ simply 
by poling with an electric field of $\sim$ 100 MV/m. If TG$_{\rm ad}$ phase is annealed 
at high-temperature, the polar T$_3$G$_{\rm pu}$  phase is formed with a TTTGTTT${\rm G}'$
conformation. In analogy to the TG phases, there might exist an non-polar T$_3$G$_{\rm au}$ 
phase as suggested by Lovinger \cite{lovinger81} and confirmed by simulations
\cite{wag92}. In fact, we will use this phase to model the non-polar phase with
T$_3$G conformation observed experimentally \cite{zhang98} in electron irradiated 
P(VDF-TrFE). The most interesting polymorph for ferroelectricity is phase T$_{\rm p}$; 
it can be obtained by mechanical drawing from T$_3$G$_{\rm pu}$ phase or poling from 
TG$_{\rm pd}$ with a $\sim$ 500 MV/m electric field. The polymer chains are in an 
all-T configuration and packed in a parallel fashion.

Changes in molecular conformation (T, TG, T$_3$G) lead to significant shape changes. As 
shown in Fig. (\ref{fig:1-stru}) the all-T crystal is longer along the chain direction 
and shorter in the directions perpendicular to it than TG and T$_3$G phases. This type 
of structural change is quite different from that in the ceramic ferroelectric compounds 
in which small ionic groups change their dipole orientation by rotation and/or displacement. 
In the polymer system the dipoles are linked together by strong covalent bonds and so that
orientational change in the dipole moments requires cooperative motion of neighboring
groups through large-scale T-G conformational changes.

\section{Methods}

The {\it ab initio} QM calculations in this paper were performed using the DFT
pseudopotential code SeqQuest \cite{seq,seq2} which uses Gaussian basis sets. For all
calculations (described in Section III) we used the Perdew, Burke, and Ernzerhof
implementation of the Generalized Gradient Approximation (GGA) \cite{PBE96}. The
SeqQuest code calculates atomic forces and the stress tensor which were used to relax
positions and cell parameters.

The characterization of many important materials properties require the simulation of
large systems (thousands of atoms) for relatively long times (nanoseconds) making QM
methods impractical. Thus we use the First Principles based force field MSXX
\cite{wag92} with MD to study energetics, nucleation energies, interface mobility,
and viscoelastic properties. The MSXX force field describes the atomic interactions
with three energy terms: i) electrostatic interactions using QM-derived charges; ii)
covalent interactions (bonds, angles, torsions, and cross terms obtained using the
Hessian biased method to fit QM vibrational frequencies; and iii) van der Waals
interactions parameterized to reproduce mechanical properties of polyethylene,
graphite, CF$_4$ and poly(tetrafluoroethylene) crystals. In order to describe the
TrFE segments we extended the MSXX FF by describing the fluorine atom in the CHF
group as the F atoms in VDF (same atom type and charge), the corresponding carbon is
treated as the CF$_2$ carbon in MSXX [denoted C$_{\rm 3VF}$ as in \cite{wag92}] with
its charge modified to maintain charge neutrality. New three- and four-body terms are
calculated using combination rules. The supplementary material contains all the force
field parameters \cite{supple}. This simple extension of the MSXX FF allows us to
characterize the effects of the presence of TrFE in PVDF.

All the simulations in this work use periodic boundary conditions, with the
z axis is oriented parallel to the polymer chain direction and the y axis in
the direction of the polarization.

\section{Energetics of various crystalline phases: DFT-GGA and MSXX force field}

A key issue to understand the electro-mechanical properties of these polymers is the
relative energies and structural changes among the various phases and the origins of
such differences. We used DFT-GGA and the MSXX force field to study the relative
energetics and structures of various crystalline phases (T, TG$_{\rm ad}$, TG$_{\rm pd}$, 
T$_3$G$_{\rm pu}$, T$_3$G$_{\rm au}$) for PVDF
and P(VDF-TrFE)(50-50) copolymer (50 mole percent VDF). Table (\ref{tab:Energies})
summarizes the energetics calculated by optimizing atomic positions and cell
parameters of various phases and isolated chains with periodic boundary conditions.
Table (\ref{tab:Energies}) also gives lengths corresponding to 4 monomers of infinite
(periodic) isolated single chains (in practical terms we performed these calculations
using three dimensional periodic cells with large (50 \AA) lattice parameters in 
the directions perpendicular to the chain direction and optimized the lattice parameter 
along the chain). Here, the crystal cohesive energy (E$_{\rm coh}$) is defined as:

\begin{equation}
{\rm E}_{\rm coh} = {\rm E}_{\rm crystal} - {\rm N} \times {\rm E}_{\rm isolated-chain},
\end{equation}

\noindent where N is the number of chains inside the crystal phase, E$_{\rm crystal}$
is the energy of the crystal phase and E$_{\rm isolated-chain}$ is the energy of
corresponding (same chain conformation) isolated chain. All energies in Table
(\ref{tab:Energies}) are given in kcal/mol per carbon atom. DFT-GGA predicts the
TG conformation as the ground state for the isolated chains. Next, in relative 
stability (0.97 kcal/mol higher) is the T$_3$G conformation while the all-T chain 
has a much higher energy (1.41 kcal/mol higher than T$_3$G). We find that adding 
fluorine to make P(VDF-TrFE) leads to substantial stabilization of the all-T 
configuration. Although the TG conformation is still most stable (by 1.08 kcal/mol), 
the T$_3$G structure is higher in energy than all-T by 0.06 kcal/mol. Thus 
the extra F in the copolymer favors all-T conformations. The MSXX force field 
correctly describes the relative stability among different conformations in PVDF 
and the role of F in P(VDF-TrFE)(50-50), in addition to a good description of the 
lengths along the chain direction for various conformations.

In the case of condensed-phase PVDF, DFT-GGA yields a large stabilization of the all-T 
crystal structure with a cohesive energy of 2.10 kcal/mol per carbon, followed by T$_3$G 
phases (E$_{\rm coh} \sim$ 1.1 kcal/mol). Finally TG phases are less favored by 
crystallization with E$_{\rm coh} \sim$ 0.3 kcal/mol. This stabilization of T conformations 
due to better packing, represented by the difference in cohesive energies, is not enough 
to reverse the relative energetic stability of the phases: our calculations indicate that
TG$_{\rm pd}$ phase is the most stable crystal followed by the TG$_{\rm ad}$ phase. 
The MSXX force field overestimates the cohesive energies, but exhibit trends from 
PVDF to copolymer in agreement with QM. This overestimation of E$_{\rm coh}$ leads 
to an incorrect ordering of the phases: MSXX predicts the all-T phase to be the most 
stable one. Adding F to make P(VDF-TrFE) (50-50) has the same effect as in the chains: 
it favors all-T conformations. QM predicts the T$_3$G$_{\rm au}$ phase to be the ground 
state, followed by T$_3$G$_{\rm pu}$) (0.13 kcal/mol higher) and all-T (0.30 kcal/mol
higher than T$_3$G$_{\rm au}$)). Again the effect of F is to favor T configurations; 
this could be an important factor in the better crystallinity (and consequently better 
piezoelectric properties) in the copolymer as compared with PVDF observed experimentally 
\cite{koga86}.

Zhang et al \cite{zhang98} observed experimentally a phase change from a polar all-T
to non-polar T$_3$G phase when P(VDF-TrFE) is irradiated with high-energy electrons. 
This treatment leads to a lower degree of crystallinity and consequently
shorter coherence length of the crystalline regions; this may lead to an increase in 
the chain-chain separation distance. Our simulations explored two extreme cases of 
chain-chain separations: perfect crystals and isolated chains; our results are
consistent with the observation in Zhang's experiments and quantify the effect of 
inter-chain cohesive energy.

\section{Phase transformation: nucleation and propagation}

Phase transitions can be induced in PVDF and its copolymer with TrFE by 
temperature, stress or an external electric field. Understanding the molecular
level mechanisms responsible for these phase changes is important for manufacturing 
and processing \cite{furukawa89} as well as to understand the excellent electromechanical 
properties of electron irradiated P(VDF-TrFE) samples \cite{zhang02}. 

X-ray diffraction, Fourier transform infrared spectroscopy (FTIR), and differential 
scanning calorimetry have been used to characterize the micro-structural changes
induced by high energy electron irradiation in P(VDF-TrFE) \cite{zhang02}. 
It was found that irradiation reduces the degree of crystallinity of the sample decreasing 
the size of the polar domains below a critical value where the crystalline 
regions transform to a non-polar combination of T and G bonds (mainly formed by 
T$_3$G segments). This nano-structural change leads to a desirable decrease in 
polarization hysteresis; however for irradiation doses higher than the optimum 
($\sim$75 Mrad) the hysteresis increases again \cite{zhang02}. This reentrant 
hysteresis is believed to be caused by a high density of cross-linking due to high 
irradiation doses \cite{zhang02}. The complex nano-structure resulting from 
irradiation shows excellent electromechanical properties such as $\sim$5$\%$ 
electrostrictive strain resulting from an electric field induced, reversible phase 
transition between a non-polar T$_3$G and the polar all-T crystal 
\cite{zhang98,zhang02,Li2003}.
An atomistic simulations that fully captures the nano-structure of the irradiated
copolymers would involve system sizes well beyond current capabilities. Thus
in this section we characterize two ``unit" mechanisms that play a key 
role in the properties of the irradiated copolymers and shed some light 
into phase transitions in PVDF and P(VDF-TrFE) in general.

Previous theoretical studies have focused on phase transitions in PVDF and its
copolymers, including the thermal induced ferroelectric transition 
\cite{abe01-3,abe01-4,abe01-5} and field induced TG$_{\rm 
ad}$-TG$_{\rm pd}$ transitions \cite{Dvey-Aharon_JAP1980,pertsev94}.
These phase transitions are generally believed to occur via a nucleation and 
growth process \cite{furukawa89}, governed by cooperative chain rotations 
and the motion of phase boundaries \cite{pertsev94}. 

Here we characterize : i) the nucleation of a kink (G bond pairs 
separated by T bonds) in a perfect T$_{\rm p}$ crystal; and ii) the 
mobility of an interface between T$_3$G$_{\rm ad}$ and T$_{\rm p}$ phases 
[the phase transition believed to be responsible for the large electrostrictive 
strain in electron irradiated P(VDF-TrFE) \cite{Li2003}].  We focus on the 
difference in behavior between PVDF and its 50-50 copolymer with TrFE.

\subsection{Nucleation of a kink in an all-T configuration}

In phase transitions originating from the all-T ferroelectric (I$_{\rm p}$) phase [such
as the ferroelectric-paraelectric (F-P PT)], the first unit process is the formation of gauche pairs separated by 
one or more trans bond (denoted as kinks). We use the MSXX FF to calculate the energy barrier and 
molecular level mechanisms associated with this process. 
To do this we start from a perfect T$_{\rm p}$ crystal and impose a 
harmonic restraint to one dihedral angle; we then minimize the total 
energy of the system (including the restraint) as the equilibrium angle of the restraint is changed from T to G in small 
steps (5$^{\circ}$). For each restraint angle ($\theta_0$) we fully relax the system 
including lattice parameters at zero applied stress. The PVDF simulation cell used for 
this calculation was built from the T$_{\rm p}$ unit cell (two monomers) replicating it 4 times in 
the x direction, 4 times in y and 20 times in z. The resulting cell contains 16 infinite 
chains (40 monomers in each simulation cell per chain) and 1920 atoms. The copolymer was 
obtained from the PVDF cell by randomly converting 50 $\%$ of the CH$_2$ groups into CHF.

\begin{figure*}
\includegraphics  [scale=0.85]      {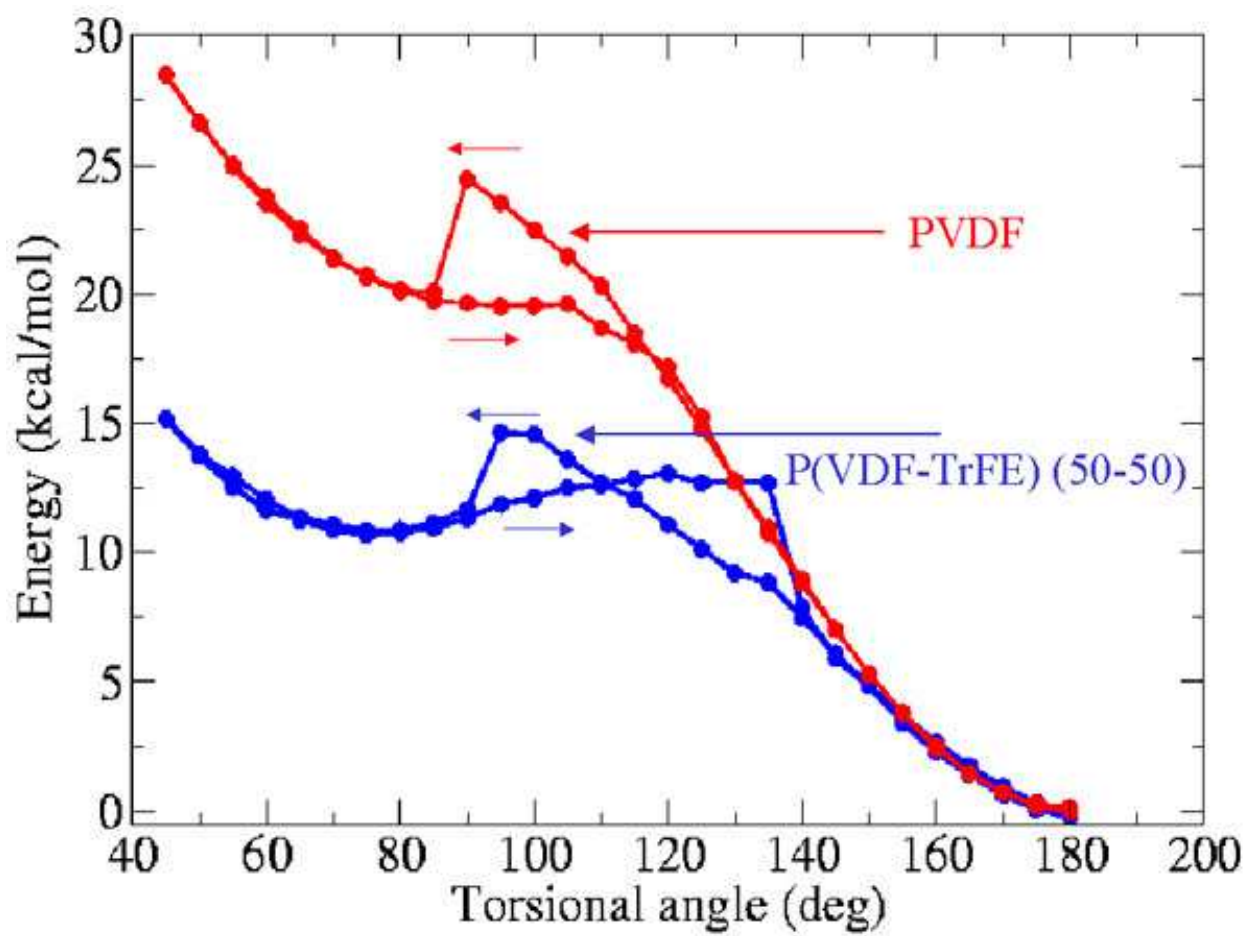}
\caption{\label{fig:nucleation} Energy barrier to nucleate a G bond in at T$_{\rm p}$ perfect
crystal both PVDF and P(VDF-TrFE)(50-50). We used supercells consisting of 16 chains 
each consisting of 8 monomers (2112 atoms). }
\end{figure*}

Fig. (\ref{fig:nucleation}) shows the total energy as a function of $\theta_0$ as the 
chosen torsion bond changes from T to G and back to T both for PVDF and P(VDF-TrFE)(50-50). 
For PVDF we obtain an energy barrier of 24.8 kcal/mol while for the copolymer the barrier 
reduces to 14.9 kcal/mol. Thus our simulations indicate that nucleation of the
F-P PT takes place much easier in P(VDF-TrFE) 
than in PVDF. Experiments show that under heating PVDF in the T$_{\rm p}$ phase melts without 
a F-P PT, but in P(VDF-TrFE) the F-P PT is observed before melting for concentrations 
of TrFE larger than 18 $\%$. While the molecular level mechanism of
the phase transition is not fully understood and defects [for instance the interface between
crystalline and amorphous regions \cite{Dvey80,pertsev94}] can play a key role, our 
calculated nucleation barriers and the experimental results on phase transitions support
nucleation and growth as the process responsible of the phase transition. 

\begin{figure*}
\includegraphics  [scale=0.4]      {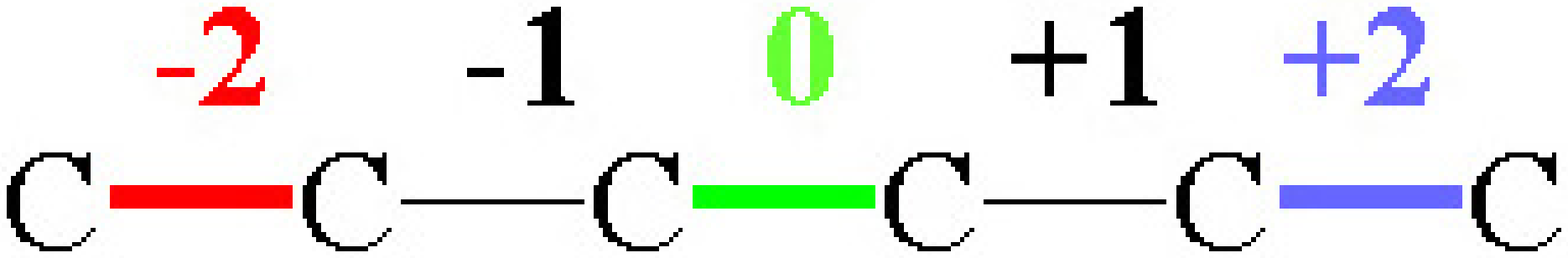}
\includegraphics  [scale=0.9]      {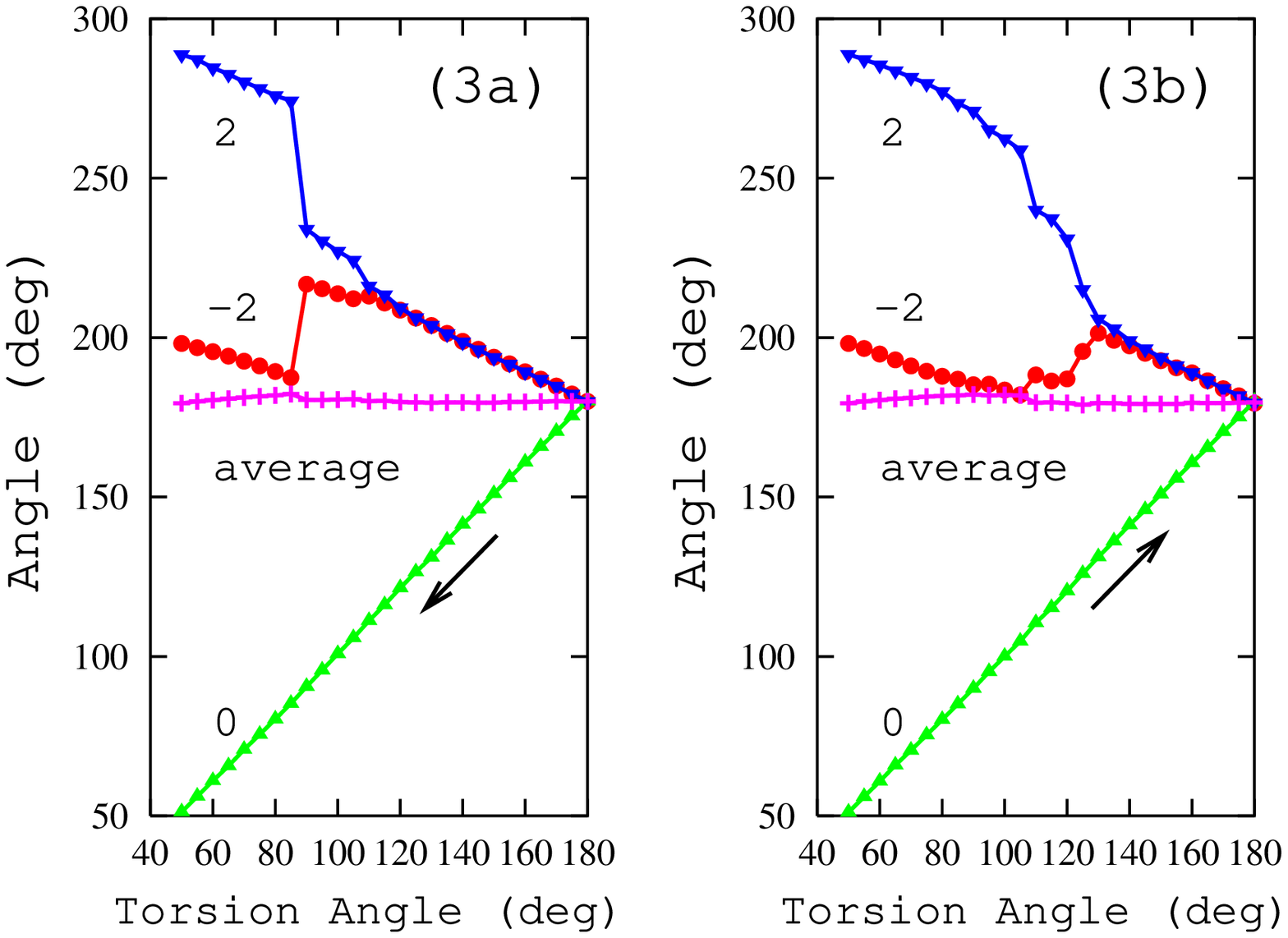}
\caption{\label{fig:cons} Evolution of various torsion angles in PVDF as a function
of the restraint dihedral angle as it is transform from T to G. The diagram at the 
top shows the notation: the restraint is applied to torsion angle for the bond denoted 
0. We find that the first nearest neighbor bonds to the restraint one  (+1 and -1) 
remain in the T conformation, while the second nearest neighbor bonds (+2 and -2)
change to compensate for the changes in bond 0. The purple curve (pluses) shows
that the average of the torsion angles remains at 180 degrees, which is topologically
necessary since the chains are infinitely periodic. Fig. (\ref{fig:cons}a) corresponds to restraint varying from 
180 to 50 while Fig. (\ref{fig:cons}b) reflects the reverse process.}
\end{figure*}

Since the polymer chains are infinitely periodic (each chain is bonded to itself at
the cell boundary) a T-to-G change in any one dihedral angle must be accompanied by 
compensating changes in the opposite direction in other bonds. To investigate this 
effect, we examined the change in dihedral angles for bonds neighboring the restrained 
one (denoted as 0) as a function of $\theta_0$, see Fig. (\ref{fig:cons}). We find 
that the angles centered around bonds that are first nearest neighbors to the restrained 
one (denoted $\pm$1) remain T, no change during the process. This is because neighboring 
G bonds are quite unfavorable energetically. We find instead that it is torsions around 
the second nearest neighbor to the restraint bonds that completely compensates for the 
change in dihedral angle of bond 0. 
Thus as bond 0 transforms to G one of the second nearest neighbors goes to $\bar{\rm G}$. The 
average between angles 0, -2, and 2 remains almost constant at the T value during the 
transformation (see magenta line in Fig. \ref{fig:cons}).

Finally, there is a correlation between the sudden energy drop seen in Fig.
\ref{fig:nucleation} and abrupt changes in the +2 and -2 dihedral angles. For small
deformations starting from the all-T conformation the +2 and -2 angles counter the
restraint in equal amounts (see Fig. \ref{fig:cons}). For $\theta_0$=90 $^o$ the
angle -2 drops back to $\sim$ 180 $^o$ and +2 doubles its contribution. This
structural relaxation leads to a drop in the energy (see Fig. \ref{fig:nucleation}).
A similar phenomena occurs when the restraint angle is changed back to 180 $^o$.

\subsection{Mobility of the interface between polar and non-polar regions}

We now turn to the second component in the nucleation and growth process: the 
velocity at which the already nucleated phase transition propagates; this is 
a critical property key to the performance of electron irradiated P(VDF-TrFE).

\begin{figure*}
\includegraphics [scale=0.85]  {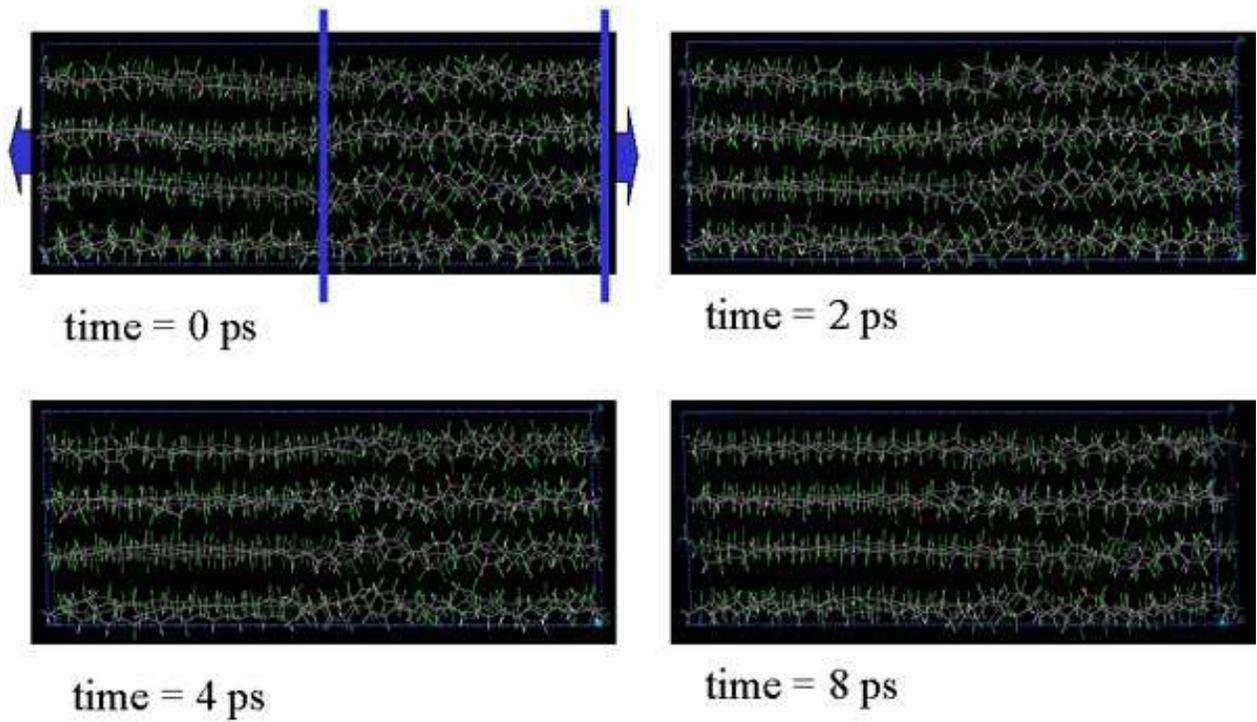}
\caption{\label{fig:3-mobility} Snapshots of the T$_3$G$_{ad}$-T$_{\rm p}$
interface for P(VDF-TrFE)(50-50) under 1.6 GPa tensile stress as various
times obtained from the MD simulation.}
\end{figure*}

Fourier transform infrared spectroscopy shows that electron irradiation of P(VDF-TrFE)
random copolymers decreases all-T content and increases T$_3$G segments \cite{zhang02};
an electric field-induced phase transition between these two phases is responsible 
for its electromechanical properties.
Therefore, we model the irradiated material by a simulation 
cell containing non-polar T$_3$G$_{\rm au}$ and all-T phases and studied the mobility 
of the interface between them under tensile stress. The interface plane is perpendicular 
to the direction of the infinite chains [see Fig. (\ref{fig:3-mobility})]. The simulations 
cell contain 2112 atoms, 4 by 4 chains in the x and y directions with each chain 
containing 44 monomers in the periodic cell. We equilibrate the system using NPT MD 
simulations with the MSXX FF at T=300 K and zero
stress. Using average lattice parameters obtained from the diagonal components of the
shape matrix during the NPT run and 90 $^\circ$ cell angles, we performed 40 ps of
NVT simulations to further equilibrate our system. After this procedure residual
stress was less than 0.05 GPa. The interface between all-T and T$_3$G
phases did not move during the equilibration process (it is well known experimentally
that drawing is necessary to obtain an all-T configuration). We then applied a
tensile uniaxial stress along the chain direction using NPT dynamics; the stress was
increased from zero at a loading rate of 0.5 GPa per ps until the desired stress was
achieved, then the applied tensile stress was maintained and the interface mobility
studied. Fig. (\ref{fig:3-mobility}) shows snapshots of this process for
P(VDF-TrFE)(50-50) with an applied stress of 1.6 GPa.

\begin{figure*}
\includegraphics  [scale=0.44]   {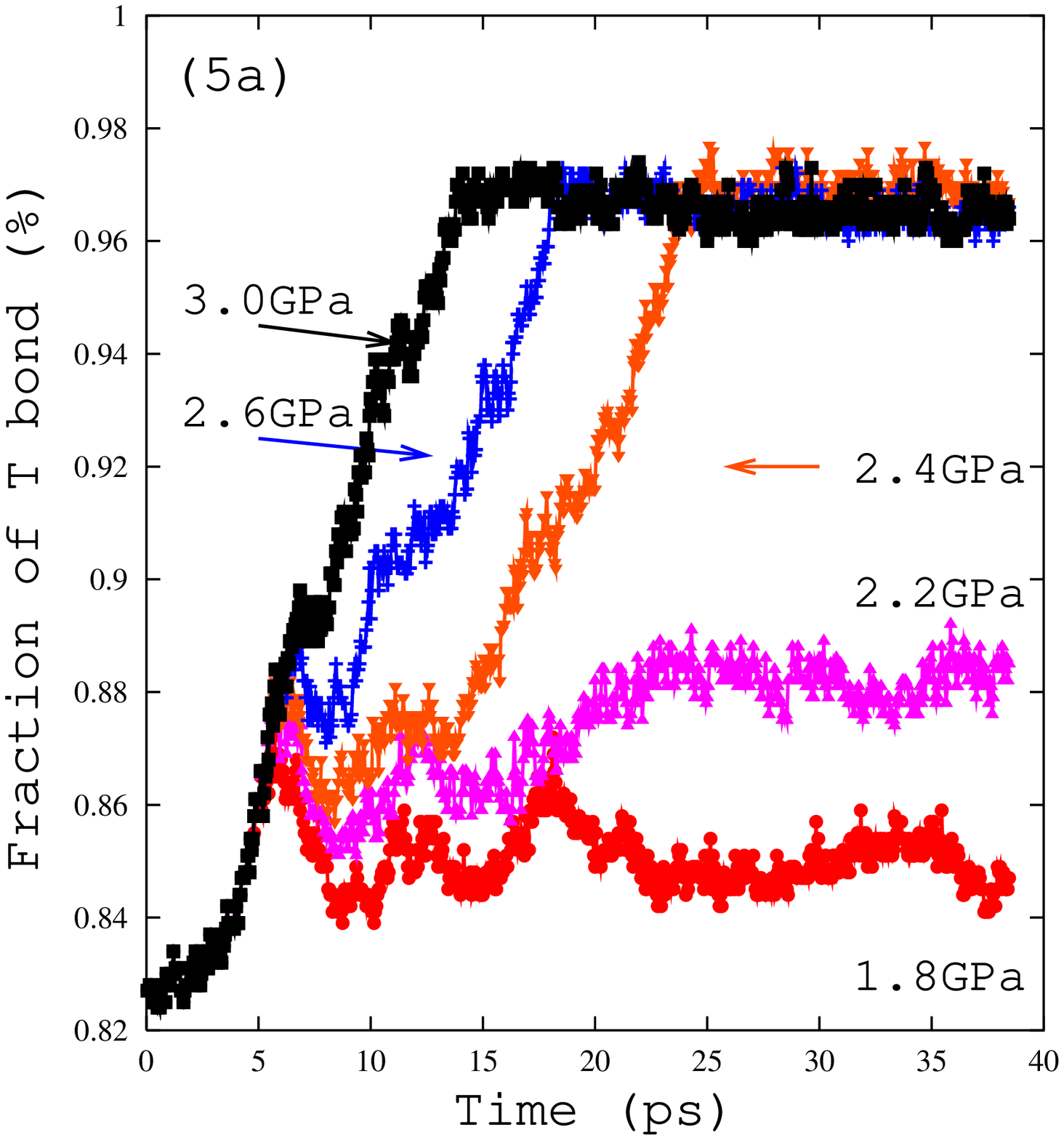}
\includegraphics  [scale=0.44]   {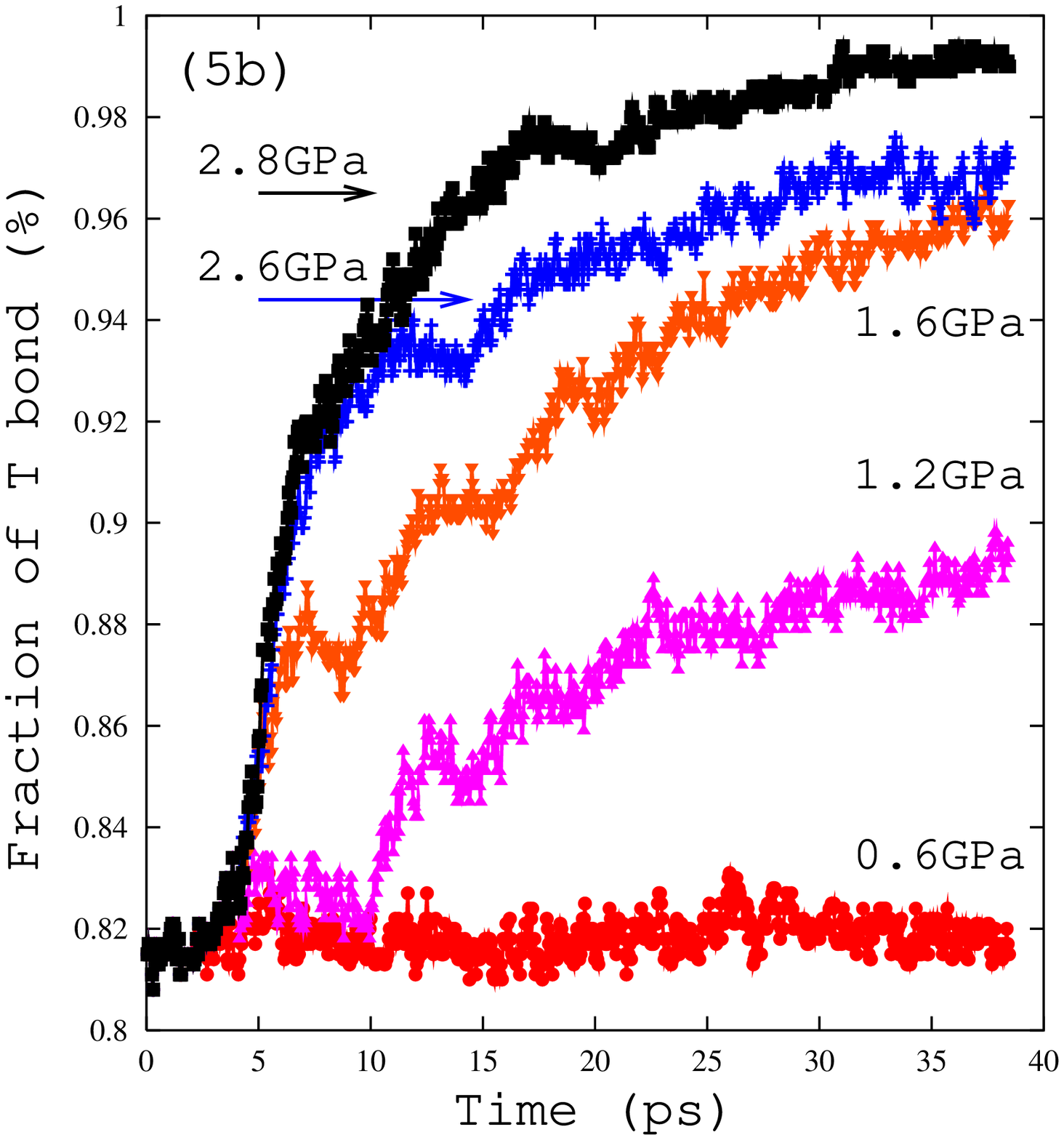}
\caption{\label{fig:3-trans} T$_3$G$_{ad}$-T$_{\rm p}$ interface mobility. Number
of T bonds as a function of time for PVDF (Fig.  (\ref{fig:3-trans}a)) and 
P(VDF-TrFE)(50-50) (Fig. (\ref{fig:3-trans}b)) under various uniaxial stress 
states at T=300 K.}
\end{figure*}

\begin{figure*}
\includegraphics  [scale=0.85]   {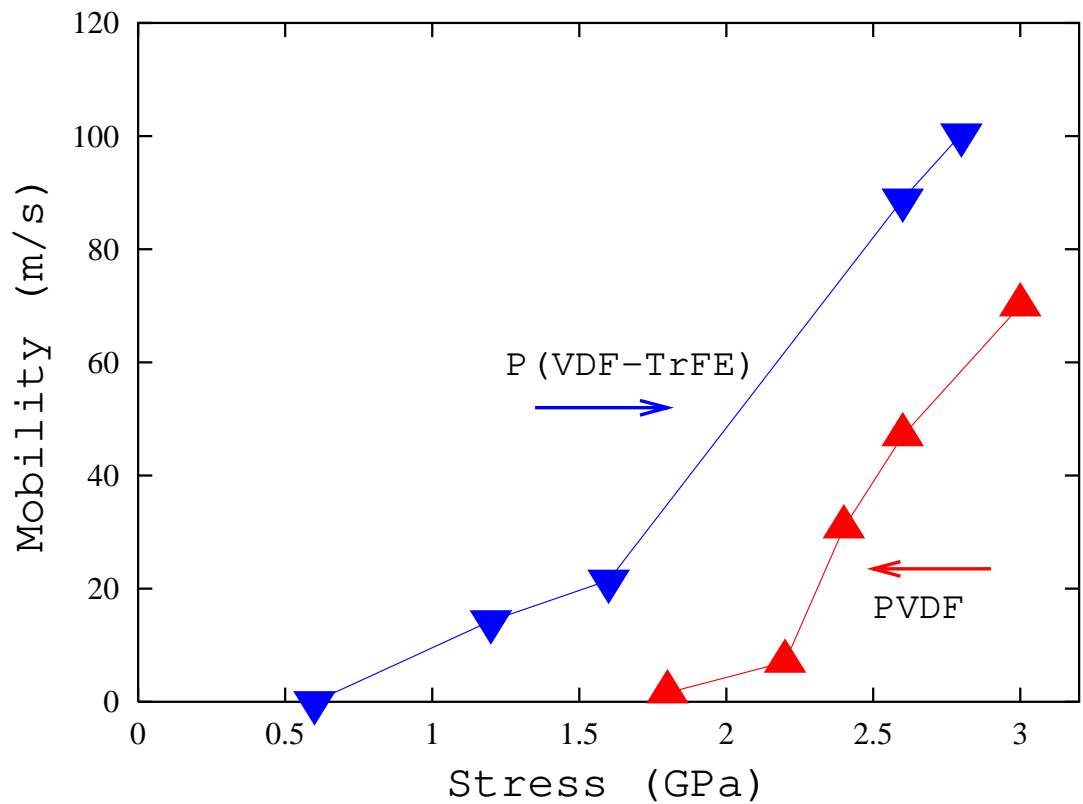}
\caption{\label{fig:3-rate} T$_3$G$_{ad}$-T$_{\rm p}$ interface mobility as a function 
of applied uniaxial stress for PVDF and P(VDF-TrFE)(50-50) at T=300 K.}
\end{figure*}

Figure \ref{fig:3-trans} shows the percentage of T bonds in the system as a function
of time for various applied stresses for PVDF \ref{fig:3-trans}(a) and P(VDF-TrFE)(50-50)
\ref{fig:3-trans} (b). Bonds are classified as T or G based on the their dihedral angles. 
From the number of G bonds as a function of time we compute interface mobility, v, using the following equation:

\begin{equation}
{\rm v} = {\delta \rm n \times \rm L \over \delta \rm t},
\end{equation}

\noindent where $\delta n$ is the change in the number of T bonds during the time period 
$\delta t$ normalized to the number chains in the simulation cell, and L is the length 
of the T$_3$G unit along chain direction. The calculated 
interface mobilities are plotted in Fig. \ref{fig:3-rate}. First, we notice that the
threshold tensile stress to move the interface for P(VDF-TrFE) is about 1 GPa, while 
for PVDF it is much larger ($\sim$ 2 GPa). This explains why it is easier to draw 
non-polar phases into polar phases in the case of the copolymer as compared with PVDF. 
Second, we find that the domain interface mobility is much larger in the case of the 
copolymer; for example, we find a velocity of 70 m/s for PVDF under a tensile stress 
of 3 GPa and 100 m/s for P(VDF-TrFE) under 2.8 GPa. These velocities are less than 
one-hundredth of the speed of sound propagating in PVDF and P(VDF-TrFE) [The sound 
velocity in the $<$001$>$ direction is about 12 km/s and 11 km/s for PVDF and 
P(VDF-TrFE) T$_{\rm p}$ phases respectively.] We note that MD simulations incorporate
in a natural way the many-body effects and cooperative phenomena involved in the
motion of a domain boundary.
     
\section{Chain sliding under shear deformation}

The various crystalline structures of PVDF and P(VDF-TrFE) are formed by covalently
bonded chains packed in a parallel fashion interacting via weak van der Waals forces. 
Since the various phases have large differences in length along the chain direction, 
we expect inter-chain slip to be an important mechanisms during the phase transition, 
especially when the interface separating the two phases is along the chains. This 
process may play an important role to release deviatoric stresses that build up during 
phase transitions [as when electrostrictive P(VDF-TrFE) is used as an actuator].

Two popular methods for studying the viscous properties in molecular simulations are:
i) the Green-Kubo (GK) method to analyze the time integral of the stress-stress
correlation function during equilibrium molecular dynamics leading
directly to the ordinary viscosity at zero strain rate \cite{cui98}, and ii)
non-equilibrium molecular dynamics (NEMD) in which a deformation is imposed on the
system and the time-averaged resulting component of the stress tensor is
calculated; the stress-strain rate data can then be used to calculate viscosity as 
a function of strain rate. In this paper we use the NEMD technique, imposing
volume-conserving pure shear deformation to our simulation cells in two directions for
various strain rates. 

\begin{figure*}
\includegraphics  [scale=0.85]   {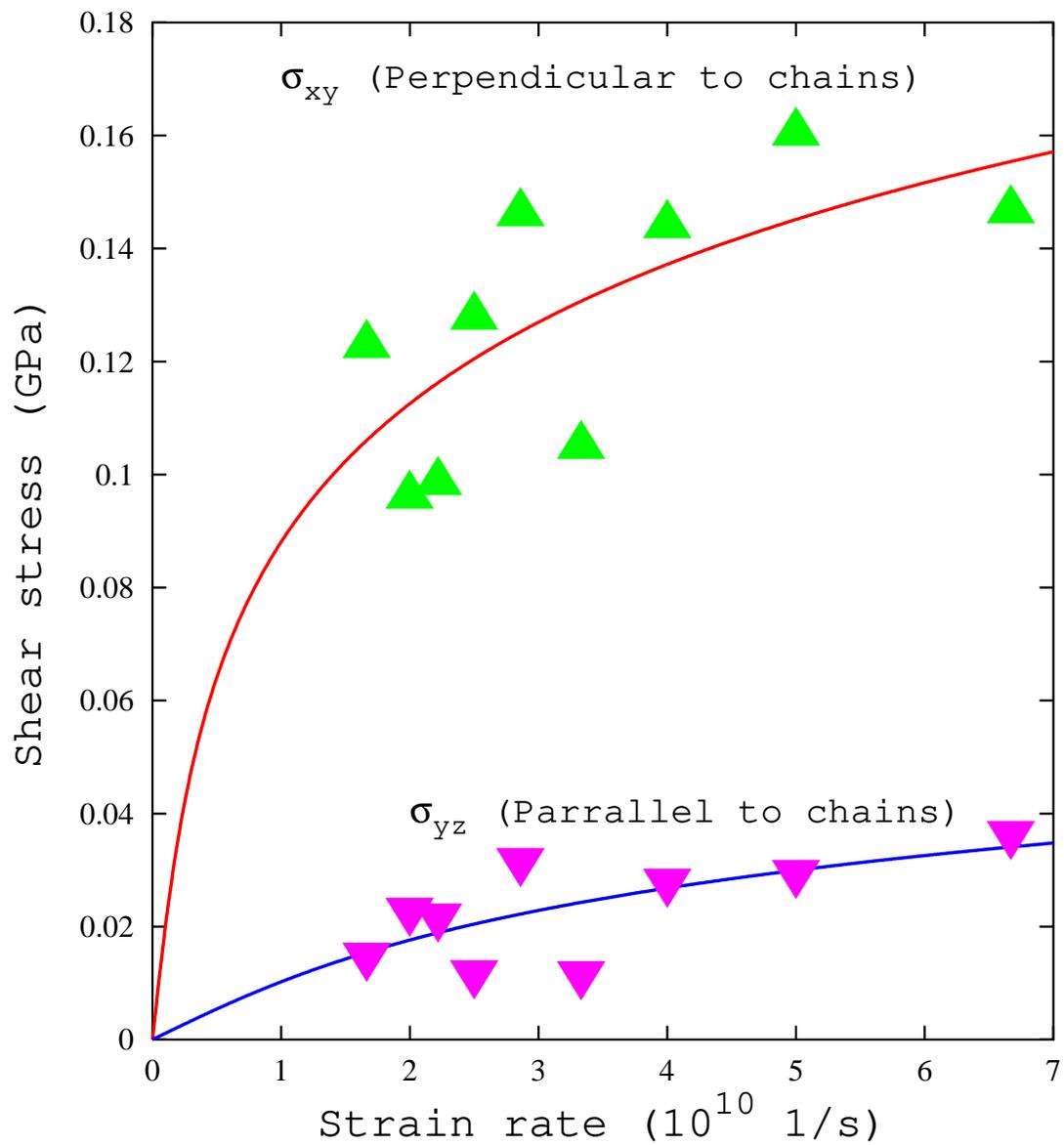}
\caption{\label{fig:3-shear}
Shear stress and a function of applied strain rate for deformation in two different 
orientations with respect to the polymer chains for P(VDF-TrFE)(50-50) 
at 300K. Lines represent fits to the MD data using the hyperbolic sine function .}
\end{figure*}

Using the convention that the c axis is the chain direction and the b axis is parallel 
to the dipole moments of the T$_{\rm p}$ phase, we deformed T$_{\rm p}$ P(VDF-TrFE) crystals along 
the following systems: i) on the (100) plane in the $\langle$010$\rangle$ direction 
(denoted as perpendicular to the chains), and ii) on the (010) plane in the
$\langle$001$\rangle$ direction (denoted as along the chains). We applied strain rates
in the range 1.66 x 10$^{10}$ 1/s to 6.67 x 10$^{10}$ 1/s using simulations cells 
containing 2 x 4 x 4 unit cells (384 atoms). Symbols in Fig. (\ref{fig:3-shear}) 
show the resulting averaged shear stress as a function of the imposed strain 
rate for both the perpendicular and parallel cases. We fitted our data to the widely 
used inverse hyperbolic sine flow equation:

\begin{equation}
\label{Eq:flow}
\sigma(\dot\gamma) = {\rm k_BT \over V_0} \rm sinh^{-1} \left( {\dot \gamma \over 
\dot\gamma_0} \right),
\end{equation}

\noindent where $\sigma$ is the shear stress, $\dot{\gamma}$ is the strain rate, 
$V_0$ denotes the effective volume of the unit deformation, and $\dot \gamma_0$ 
is a characteristic frequency (inverse time). The fitted functions are shown as 
solid lines in Fig. (\ref{fig:3-shear}). The flow equation [Eq. (\ref{Eq:flow})] 
is very general in nature and can describe various mechanisms such as free 
volume-based deformation in metallic glasses \cite{spaepen77, argon79} and 
dislocation-based single crystal plasticity \cite{stainer02, Cuitino02}. The 
resulting effective volume in the parallel case is $V_0=281.4 \AA^3$ and in the 
perpendicular direction is 116.3 $\AA^3$; the characteristic frequencies
$\dot \gamma_0$ are: 1.33 x $10^{10} 1/s$ and 0.17 x $10^{10} 1/s$ in the parallel 
and perpendicular directions respectively. 
As expected, we find that the direction of easy slip has a larger characteristic frequency
[$\dot \gamma_0$ is proportional to $\exp{(\Delta G/k_bT)}$ where $\Delta G$ is an 
activation barrier \cite{spaepen77, argon79, stainer02} ] and a larger effective 
volume. 

\begin{figure*}
\includegraphics  [scale=0.85]   {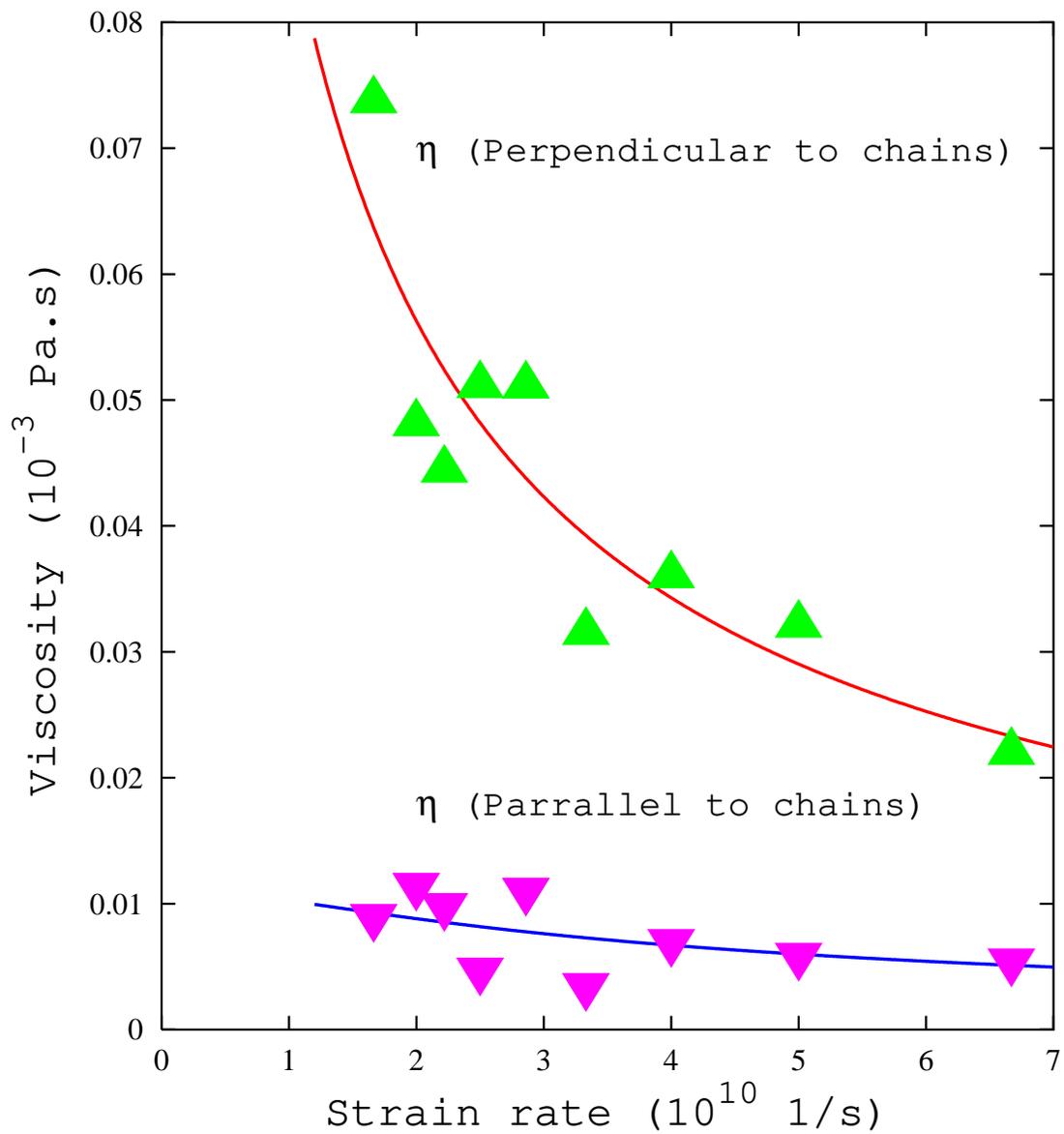}
\caption{\label{fig:3-visc}
Viscosity for two different orientations with respect to the chains of P(VDF-TrFE)(50-50) 
at 300K. Points represent MD data and lines were obtained from the fits to the stress-strain
rate data using Eq. 3.}
\end{figure*}

We calculate viscosity ($\eta$) as  $\eta = \sigma / \dot{\gamma}$;
Figure (\ref{fig:3-visc}) shows the viscosity as a function of strain rate for both 
directions. The symbols represent the atomistic data and the lines were obtained from 
the previous fits. We see from Fig. (\ref{fig:3-visc}) that we are in the shear-thinning 
regime (an increase in the shear rate leads to a decrease in viscosity) and the 
viscosity along the chains direction (the one important to release stress during a 
phase transition) is significantly smaller than that in the perpendicular direction. 
This is somewhat evident by examining the T$_{\rm p}$ crystal structure and more clear by 
analyzing the MD trajectories. Fig. (9) shows snapshots from our runs for 
$\dot \gamma = 4.0$ x $10^{10}$ both in the parallel (top panels) and perpendicular 
(bottom panels) directions. While both deformations lead to disorder of the chains, 
the perpendicular case involves much larger molecular rotations and relative shifts, 
and consequently larger energies and stresses.

\section{Conclusions}

We used a combination of {\it ab initio} (DFT-GGA) calculations and MD
with First Principles Force Fields to characterize several ``unit mechanisms"
that govern the electromechanical properties of electroactive polymers PVDF
and P(VDF-TrFE)(50-50). The quantum mechanics calculations of energetics and 
structures of various phases of PVDF and P(VDF-TrFE) show that T configurations 
are stabilized energetically by the addition of bulkier F atoms in TrFE.
This result is likely related with the better crystallinity and piezoelectric 
properties of P(VDF-TrFE), because of its stabilization of the all-T form, T$_{\rm p}$.
Our results also show that conformations with T and G bonds are energetically
favorable for large inter-chain separations as compared with all-T structures; 
this observation is consistent with the experimental observation that samples 
bombarded with high-energy electrons favor T$_3$G non-polar conformations.

Our MD results show that the energy barrier required to nucleate a G bond in an 
all-T configuration in P(VDF-TrFE) is $\sim$ 40$\%$ lower than in PVDF. 
Experimentally copolymers show a P-N PT upon heating 
before melting while PVDF directly melts; this is consistent with the calculated 
nucleation barriers.

We also studied the motion of the interface separating non-polar T$_3$G$_{au}$
and polar T$_{\rm p}$ phases under uniaxial stress to model the reversible phase transition
responsible for electrostriction in electron irradiated P(VDF-TrFE). We found a 
smaller threshold stress and higher domain wall mobility in P(VDF-TrFE) 
as compared to PVDF. Interface mobility is an important property for actuators 
and MD provides a detailed understanding and quantitative characterization of this
process that should be useful in meso- or macro-scopic simulations.
Finally, our simulations of the dynamics of chain sliding reveal that 
slip occurs more easily along the direction of the chains than perpendicular to
them.

In  summary we have used First Principles-based multi-scale modeling to characterize
dynamical and static properties PVDF and its copolymers with TrFE. Our simulations
are consistent with available experimental observations. In order to make a more quantitative
comparison between our dynamical calculations and experiments, we are currently using 
the atomistic results with mesoscopic modeling \cite{Cuitino2003} to {\it predict} 
the behavior of a real actuator whose performance can be directly compared to experiments.
We foresee that such multi-scale simulations will be an important tool to guide the design
of new materials with improved properties.

\begin{acknowledgments}

The work has been funded by DARPA and ONR (Program Managers Carey Schwartz and Judah 
Goldwasser). We thank A. Cuiti\~ no for many fruitful discussions.
The facilities of MSC used in these calculations were supported by ONR-DURIP, ARO-DURIP, 
NSF-MRI, and IBM-SUR. In addition the MSC is supported by NSF, NIH, ONR, General Motors, 
ChevronTexaco, Seiko-Epson, Beckman Institute, Asahi Kasei, and Toray.

\end{acknowledgments}



\newpage

\begin{table*}
\caption{Energetics of crystalline phases and isolated infinite chains of PVDF and
P(VDF-TrFE)(50-50). All energies are given in kcal/mol per carbon atom. The reference
energies for crystal phases and chains are the energy of T$_{\rm p}$ phase and all-T chain,
respectively. The cohesive energy defined by Eq. (1) in the text is denoted by
E$_{\rm coh}$. The length (L) of isolated chain is defined as the length of four
monomers along the chain direction. (QM) denotes DFT-GGA data;
(FF) denotes MSXX force
field.} \label{tab:Energies}
\begin{tabular}{|l|c|c|c|c||c|c|c|c|}
\colrule
        & \multicolumn{4}{c||}{\bf PVDF}            &  \multicolumn{4}{c|}{\bf P(VDF-TrFE)}    \\
\colrule
 Crystal&$\Delta$E (QM) & E$_{\rm coh}$(QM) &$\Delta$E (FF) &E$_{\rm coh}$ (FF) & $\Delta$E (QM) & E$_{\rm coh}$(QM) &$\Delta$E (FF) &E$_{\rm
coh}$ (FF)   \\
\colrule
T$_{\rm p}$                 &   0           &-2.10       &0            &-3.98       &0             
&-0.86       &0             &-3.02 \\
TG$_{\rm ad}$        &-0.59          &-0.31       &0.39         &-2.40       &1.86          
&2.09        
&1.17          &-0.72 \\
TG$_{\rm pd}$        &-0.62          &-0.34       &0.35         &-2.45       &2.42          
&2.65        
&1.26          &-0.62 \\
T$_3$G$_{\rm pu}$   &-0.51          &-1.20       &0.19         &-3.14       &-0.17         
&-1.10  &0.58          &-2.49\\
T$_3$G$_{\rm au}$   &-0.38          &-1.03       &0.23         &-3.10       &-0.30         
&-1.23  &0.56          &-2.50\\
\colrule
 Single Chains&$\Delta$E (QM) & $\Delta$E (FF) & L ($\AA$) (QM) & L ($\AA$) (FF) & $\Delta$E (QM) & $\Delta$E (FF)
&L ($\AA$) (QM) & L ($\AA$) (FF) \\
\colrule
T                 &    0 &  0    &   5.22  & 5.13 &   0     &  0    & 5.19 & 5.21    \\
TG        &-2.38 & -1.19 &   4.66  & 4.56 &  -1.08  & -1.13 & 4.71 & 4.68    \\
T$_3$G &-1.41 & -0.65 &   4.62  & 4.54 &   0.06  &  0.05 & 4.72 & 4.70    \\
\colrule
\end{tabular}
\end{table*}

\newpage

\begin{figure*}
\includegraphics  [scale=0.85]   {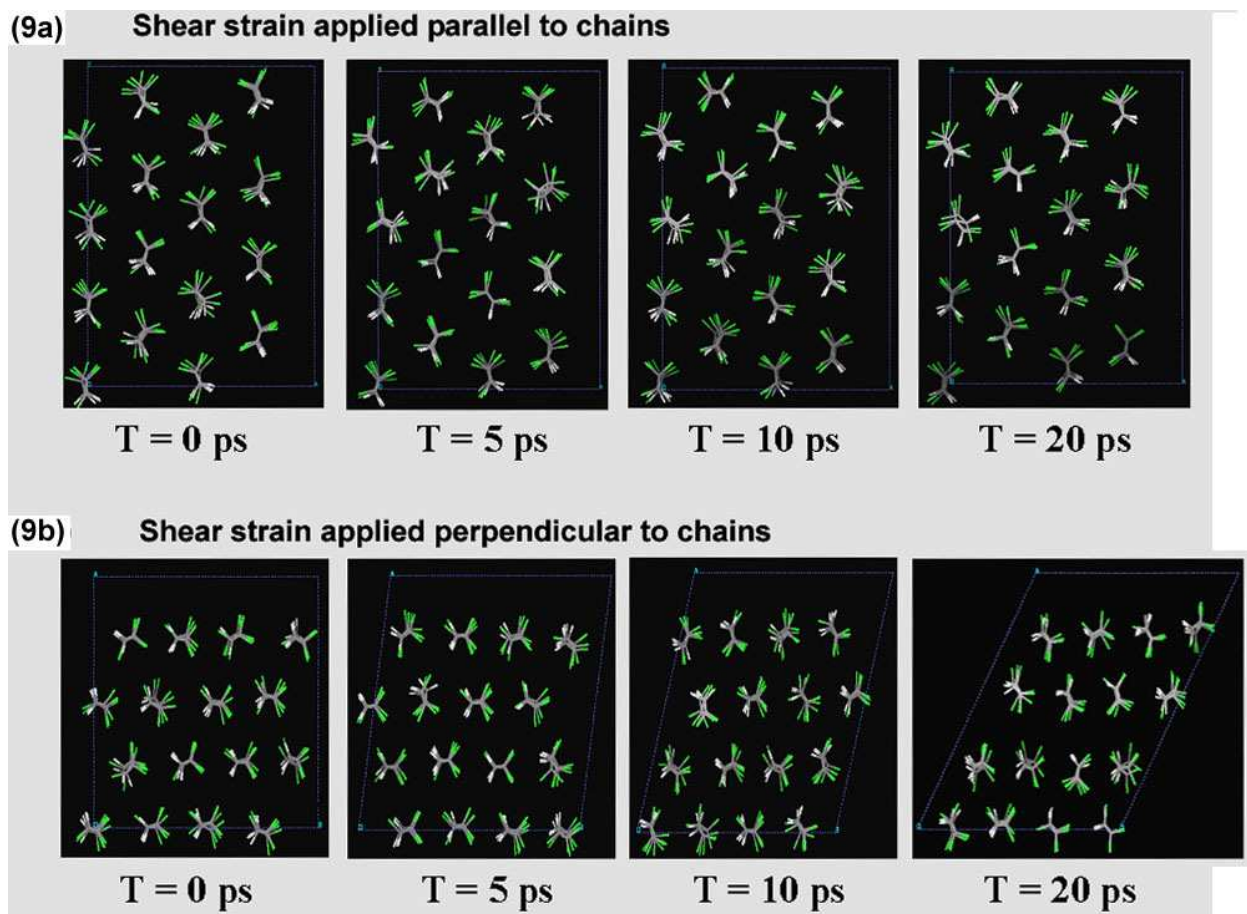}
\caption{\label{fig:3-snap}
Snapshots obtained form MD simulation of shear deformation of a P(VDF-TrFE) (50-50) crystal
using strain rate of 4.0 x 10$^{10}$ 1/s in the direction parallel (a) and 
perpendicular (b) to the polymer chains. 
}
\end{figure*}

\end{document}